%% file: main.tex
\documentclass[twocolumn]{aastex701}

\usepackage{aas_macros}
\usepackage{graphicx}
\graphicspath{{./}{figures/}}
\usepackage{dcolumn}
\usepackage{bm}
\usepackage[usenames,dvipsnames]{xcolor}
\hypersetup{
	colorlinks=true,
	citecolor=oxblood,
	linkcolor=oxblood,
	urlcolor=oxblood,
}
\usepackage{epsfig}
\usepackage{url}
\usepackage[normalem]{ulem}
\usepackage{latexsym}
\usepackage{epsfig}
\usepackage{amsmath}
\usepackage{amssymb}
\usepackage{wasysym}
\usepackage{graphicx}
\usepackage{verbatim}
\usepackage{enumerate,mdwlist}
\usepackage[titletoc]{appendix}
\usepackage{amsfonts}
\usepackage{pgfplots}
\usepackage[export]{adjustbox}
\usepackage{bbold}

\definecolor{oxblood}{rgb}{0.5333, 0.0314, 0.0314}

\input{new_commands}

\input{macros}
\usepackage[acronym]{glossaries}
\newacronym{GW}{GW}{gravitational wave}
\newacronym{EM}{EM}{electromagnetic} 
\newacronym{GL}{GL}{gravitational lensing}
\newacronym{GO}{GO}{geometric optics} 
\newacronym{WO}{WO}{wave optics}
\newacronym{CBC}{CBC}{compact binary coalescence}
\newacronym{BBH}{BBH}{binary black hole}
\newacronym{BNS}{BNS}{binary neutron star}
\newacronym{NSBH}{NSBH}{neutron-star black-hole binary}
\newacronym{PBH}{PBH}{primordial black hole}
\newacronym{SIS}{SIS}{singular isothermal sphere}
\newacronym{SIE}{SIE}{singular isothermal ellipsoid}
\newacronym{SNR}{SNR}{signal-to-noise ratio}
\newacronym{PE}{PE}{parameter estimation}
\newacronym{LVK}{LVK}{LIGO--Virgo--KAGRA}
\newacronym{IMBH}{IMBH}{intermediate-mass black hole}
\newacronym{GC}{GC}{globular cluster}

\makeatletter
\newcommand*{\glsplainhyperlink}[2]{%
  \colorlet{currenttext}{.}%
  \colorlet{currentlink}{\@linkcolor}%
  \hypersetup{linkcolor=currenttext}%
  \hyperlink{#1}{#2}%
  \hypersetup{linkcolor=currentlink}%
}
\let\@glslink\glsplainhyperlink
\makeatother

\begin{document}

\title{Probability of gravitational-wave lensing by intermediate-mass black holes and globular clusters}

\author[orcid=0000-0001-7697-8361,sname='Vujeva']{Luka Vujeva}
\affiliation{Center of Gravity, Niels Bohr Institute, Blegdamsvej 17, 2100 Copenhagen, Denmark}
\email[show]{\href{mailto:luka.vujeva@nbi.ku.dk}{luka.vujeva@nbi.ku.dk}}

\author[orcid=0000-0002-7213-3211,sname='Ezquiaga']{Jose Mar\'ia Ezquiaga
}
\affiliation{Center of Gravity, Niels Bohr Institute, Blegdamsvej 17, 2100 Copenhagen, Denmark}
\email{jose.ezquiaga@nbi.ku.dk}
\author[orcid=0000-0003-1561-6716,sname='Lo']{Rico K.~L.~Lo 
}
\affiliation{Center of Gravity, Niels Bohr Institute, Blegdamsvej 17, 2100 Copenhagen, Denmark}
\email{kalok.lo@nbi.ku.dk}

\author[orcid=0000-0003-4818-3400,sname='Zwick']{Lorenz Zwick}
\email{lorenz.zwick@nbi.ku.dk}
\affiliation{Center of Gravity, Niels Bohr Institute, Blegdamsvej 17, 2100 Copenhagen, Denmark}

\begin{abstract}
Strongly lensed \glspl{GW} are powerful probes of substructure in the lens. 
\Glspl{IMBH} are postulated to be efficient lenses that may distort the lensed waveforms of currently detectable stellar-mass compact binaries, as hinted by GW231123. 
Assuming that \glspl{IMBH} are located in \glspl{GC}, we compute the rate at which they would affect strongly lensed repeated chirps produced by galaxy-scale lenses, considering a compound lens system. 
Exploring different astrophysical model assumptions and lensing criteria, we find that the relative rate is at most 1/1000 and decays to 1/10,000 for our fiducial optimistic scenario. 
At high magnifications, $\mu>100$, the relative rate approaches 1\%, but these cases are intrinsically rare in absolute value.  
Our results imply that \gls{GW} lensing by \glspl{IMBH} and \glspl{GC} is unlikely,
disfavoring such an interpretation for GW231123.
In turn, they point towards lensed \glspl{GW} being a clean probe of dark matter substructures and primordial black holes.  
\end{abstract}

\glsresetall
\section{Introduction}

At design sensitivity of current ground-based detectors, 1--10 strongly lensed \glspl{GW} are expected every year~\citep{Xu:2021bfn,Wierda:2021upe,Smith:2022vbp,Li:2026dai}. 
These lensed \gls{GW} signals promise to have a broad scientific impact. 
With just a few observations, they will enable unprecedented tests of additional polarizations~\citep{Goyal:2020bkm}, 
dark-energy theories~\citep{Ezquiaga:2020dao,Goyal:2023uvm}, and \gls{GW} propagation~\citep{Finke:2021znb,Chung:2021rcu}. 
Moreover, they will significantly improve sky localization for dark siren cosmography~\citep{Hannuksela:2020xor,Wempe:2022zlk}. 
Similarly, they could provide clues to resolve astrophysical \gls{BBH} environments~\citep{Samsing:2024xlo,Samsing:2025rxq} and different formation histories~\citep{Ng:2020qpk,Ezquiaga:2020tns,Franciolini:2024vis}. 
While searches for signatures of gravitational lensing of \glspl{GW} are being conducted with data from the LIGO \citep{LIGOScientific:2014pky}, Virgo \citep{VIRGO:2014yos} and KAGRA \citep{KAGRA:2020tym} detectors, there is no conclusive evidence yet \citep{Hannuksela:2019kle,LIGOScientific:2021izm,Janquart:2023mvf,LIGOScientific:2023bwz,GWTC4_lensing}.

\gls{GW} detectors have exquisite time resolutions -- down to the millisecond -- becoming powerful probes of any cosmic (sub)structure. 
This is particularly relevant to probe the nature of dark matter at sub-galactic scales, either if in the form of compact objects~\citep{Diego:2019rzc} or subhalos~\citep{Oguri:2005je,Dai:2018mxx,Vujeva:2025nwg, Ando:2026poq, Ando:2026eam}. 
Interestingly, the time delays associated with these substructures can be comparable to the duration of the \glspl{GW} themselves, making the potential repeated chirps produced by strong lensing interfere with each other and even capture frequency-dependent distortions due to wave-optics lensing~\citep{Nakamura:1997sw}. 

\Glspl{IMBH} may also be efficient lenses for \glspl{GW}~\citep{Lai:2018rto,Meena:2023qdq}. 
In fact, recent \gls{GW} observations have brought a puzzling lensed candidate, GW231123~\citep{GW231123,GWTC4_lensing}.  
The interpretation of the signal is consistent with a $\sim1000M_\odot$ point-lens mass embedded in a galactic halo~\citep{Goyal:2025eqo}. 
Although large-scale simulation campaigns show that the statistical significance of the lensing hypothesis is bounded to $4\sigma$~\citep[lowered to $\sim3\sigma$ when accounting for potential waveform systematics]{Chan:2025kyu}, an estimate of the astrophysical likelihood of such a class of configuration is still missing. 
A discovery of a \gls{GW} lensed by an \gls{IMBH} would have profound consequences, as 
observations of isolated black holes in this mass range are challenging.  
Non-\gls{GW} lensed transient candidates have been identified~\citep{Paynter:2021wmb,Levan:2025ool}, but their evidence has been contested~\citep{Mukherjee:2023qkg}.

\begin{figure*}[t!]
    \centering
    \includegraphics[width=\linewidth]{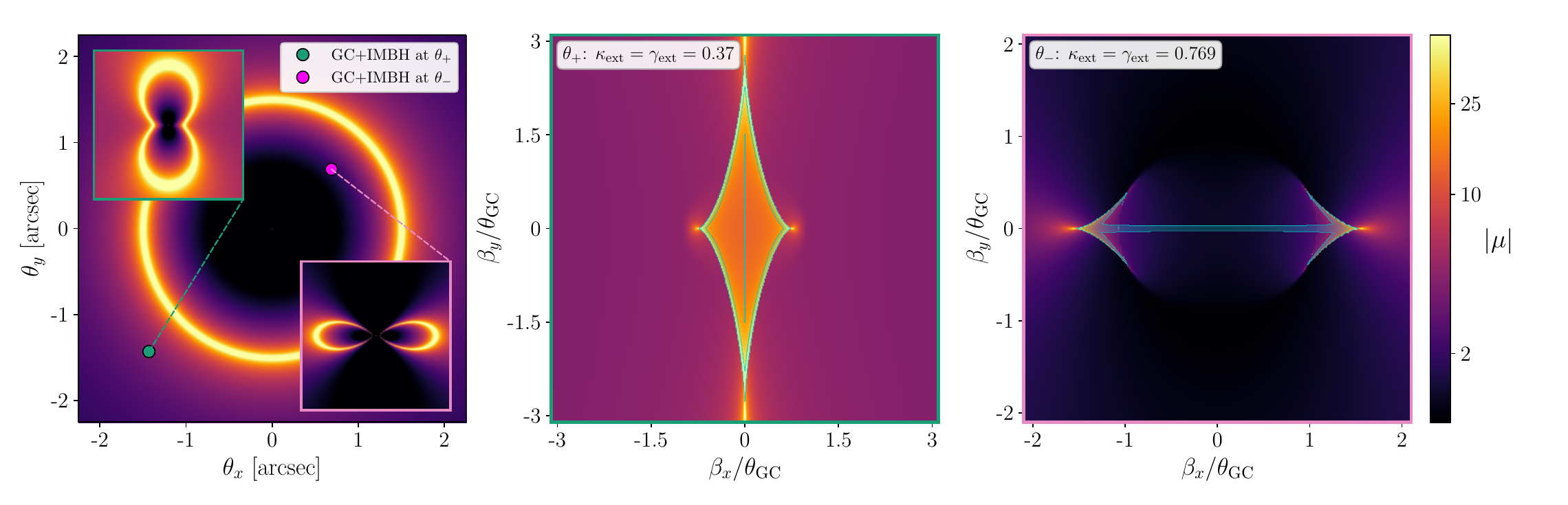}
    \caption{Schematic view of the problem. 
    A \gls{GW} is strongly lensed by a galaxy producing repeated chirps (colored circles in the image plane at arcsecond separations, left panel). 
    We study the probability that one of the repeated copies is further distorted by the lensing produced by an \gls{IMBH} embedded in a GC
    and subject to the external convergence of the main halo (inset plots at hundreds of micro-arcseconds). 
    The cyan shaded region corresponds to the area in the source plane (middle and right panels) that satisfies our lensing criterion -- time delays between the brightest images less than 1 second and relative magnifications larger than 0.1. 
    The lensing cross sections are noticeably different for the embedded lens outside ($\theta_+$) or inside ($\theta_-$) the critical curve due to the presence of the \gls{IMBH}. 
    The color map indicates the absolute value of the magnification. 
    }
    \label{fig:the_problem}
\end{figure*}

Other astrophysical structures within the wave-optics regime of current \gls{LVK} sources are \glspl{GC}. 
They have been considered as potential lenses of continuous \glspl{GW} in our own galaxy~\citep{Moylan:2007fi,Suvorov:2021uvd}, although the inferred rates are at the level of $\sim10^{-4}$. 
\glspl{GC} in the Milky Way have also been studied as potential lenses of cosmological \glspl{BBH}, finding that the \gls{GC} velocity dispersion could be accurately reconstructed~\citep{Harikumar:2026pnu}. 
However, \gls{GC} lensing within our galaxy for cosmological sources is also intrinsically rare, with optical depths at the level of $\sim10^{-8}$~\citep{Harikumar:2026pnu}.

\section{Modeling the population of astrophysical intermediate-mass black holes and globular clusters}

\Glspl{IMBH} are the subject of extensive observational searches and theoretical modeling efforts~\citep{Greene:2019vlv}. Their formation channels, abundance, and population properties remain poorly constrained. Broadly, three astrophysical \gls{IMBH} populations have been proposed. First, \glspl{IMBH} may reside in the nuclei of dwarf galaxies~\citep{2018MNRAS.478.2576M,2023MNRAS.523.5610B}. Second, \glspl{IMBH} may populate the halos of larger galaxies as relics of past minor mergers and tidal stripping events~\citep{2005MNRAS.358..913V,2023MNRAS.525.1479D}. Third, \glspl{IMBH} may form and persist in dense stellar systems, in particular  \glspl{GC} , through runaway stellar collisions, repeated black hole mergers, or other dynamical processes~\citep{2002ApJ...576..899P,2002MNRAS.330..232C}.

In this work we focus on the latter scenario, namely \glspl{IMBH} associated with  \glspl{GC}, as the first two populations are either negligible or too unconstrained to incorporate into a rate estimate. Central \glspl{IMBH} in dwarf galaxies contribute only a single compact lens per satellite galaxy, whose lensing cross section is typically much smaller than that of the dwarf galaxy itself. The abundance and spatial distribution of wandering \glspl{IMBH} produced by tidal stripping and minor mergers is observationally unconstrained. Theoretical modeling claims reach numbers of up to 10 to 100 \glspl{IMBH} per halo at high redshifts \citep{2023MNRAS.525.1479D,2025MNRAS.538.2255V}, though their distribution is shaped by complex dynamical processes in galaxies \citep{2022MNRAS.510..531C} that are still unconstrained at these mass scales \citep{LISA:2017pwj,2024arXiv240207571C}.

In contrast,  \glspl{GC} constitute a visible tracer population with well-measured abundances, masses, and spatial distributions within galaxy
halos. They therefore provide an astrophysically motivated reference population for estimating the probability that a strongly lensed \gls{GW} undergoes additional lensing by an \gls{IMBH}. Furthermore, \glspl{GC} are typically found at projected distances of $\sim10$--$100\,{\rm kpc}$ from the halo center, comparable to the Einstein radii of galaxy-scale strong lenses, making them natural candidates for producing additional lensing.
Finally, massive galaxies commonly host $\mathcal{O}(10^3)$  \glspl{GC}, suggesting that \glspl{IMBH} residing in  \glspl{GC} may constitute the dominant population of intermediate-redshift compact lenses contributing to the lensing kernel. We therefore adopt this population as our fiducial model for \gls{IMBH} lensing.

Given these considerations, our goal is to estimate the probability that a strongly lensed \gls{GW} is also lensed by an \gls{IMBH}/\gls{GC} within the halo.  The key ingredients are the spatial distribution of \glspl{IMBH}, their masses, and their abundance. We parameterize the occupation fraction of \glspl{IMBH} in  \glspl{GC}  by:
\begin{align}
    N_\bullet \equiv f_{\rm occ}\,N_{\rm GC},
\end{align}
where $N_\bullet$ is the number of \glspl{IMBH} and $N_{\rm GC}$ the total number of  \glspl{GC}  within the halo. 
We further parameterize the \gls{IMBH} mass as a fixed fraction of the host globular cluster mass:
\begin{align}
    M_\bullet \equiv f_{\rm mass}\,M_{\rm GC},
\end{align}
where $f_{\rm mass}\leq1$. 
The parameter $f_{\rm mass}$ determines 
whether the lensing is dominated by the \gls{GC} or by the central \gls{IMBH}, which can be inferred by comparing their relative Einstein radii.

Given these simple parameterizations, we now summarize the observed properties of \gls{GC} systems. Our goal is to define a representative astrophysical population against which the lensing probabilities can be estimated, not to provide a detailed overview. Therefore, we will use the simplest available scaling laws and order of magnitude estimates, anchored to empirical scaling relations measured for \gls{GC} systems in nearby galaxies. For Milky Way-like galaxies, the total mass in  \glspl{GC}  is approximately~\citep{2014ApJ...787L...5H,2018MNRAS.477.3869H}
\begin{align}
    M_{\rm GC}^{\rm tot}\simeq3\times10^{-5}\,M_{\rm h},
\end{align}
where $M_{\rm h}$ denotes the halo mass. The total number of  \glspl{GC}  scales approximately as~\citep{2018MNRAS.477.3869H,2020AJ....159...56B}
\begin{align}
    N_{\rm GC}\simeq2\times10^{-10}\,
    \frac{M_{\rm h}}{M_\odot}.
\end{align}
For a representative halo mass of $M_{\rm h}\sim10^{13}\,M_\odot$, this gives a population of approximately $2\times10^3$  \glspl{GC} . As a consistency check, scaling the relation down to a Milky Way-sized halo with $M_{\rm h}\lesssim10^{12}\,M_\odot$ \citep{2019ApJ...871..120E,2020MNRAS.494.4291C} predicts $\lesssim200$  \glspl{GC} , in good agreement with the $\sim168$ confirmed  \glspl{GC}  observed in the Milky Way \citep{2019MNRAS.482.5138B}.
Combining these two relations gives a characteristic \gls{GC} mass:
\begin{align}
    M_{\rm GC}\simeq1.5\times10^5\,M_\odot,
\end{align}
which is approximately independent of halo mass. In reality, \gls{GC} masses follow log-normal distributions with means of roughly the same scale and a standard deviation of $\sim 0.5$ dex \citep{2007ApJS..171..101J}. These properties are also expected to evolve with redshift \citep{2015MNRAS.454.1658K}, an effect that we omit due to the lensing efficiency peaking in a relatively narrow range of $z_\mathrm{L} \sim $ 0.3--0.5 for the source redshifts considered in this work~\citep{Robertson:2020mfh}.

The physical size of a \gls{GC} is most commonly characterized observationally by its half-light radius. Converting this quantity into intrinsic structural properties, such as the half-mass or virial radius, requires assuming a density profile, typically a King model or a Plummer sphere \citep{Plummer:1911zza,King:1966fn}. For simplicity, however, we model  \glspl{GC}  as truncated \gls{SIS}, which allows for a straightforward treatment of the lensing calculations while remaining broadly consistent with their observed structural properties. Observed half-light radii range from a few to several tens of parsecs \citep{1995AJ....109..218T}, with the Milky Way population clustering around $\sim3\,\mathrm{pc}$ \citep{1995AJ....109..218T,baum}. We therefore adopt a half-light radius of $3\,\mathrm{pc}$ and approximate it as the half-mass radius. For a truncated \gls{SIS}, this corresponds to a virial radius of $\sim6\,\mathrm{pc}$. Combined with the adopted cluster mass, this implies a characteristic velocity dispersion of $\sim10\,\mathrm{km\,s^{-1}}$, in good agreement with observations \citep{1976ApJ...204...73I}. Although simplified, this model captures the relevant physical scales while considerably simplifying the lensing calculations.

\Glspl{GC} are distributed over galactic halo scales. Modern observational studies describe the projected surface density of \gls{GC} systems using Sérsic profiles \citep{1963sersic,1999ciotti},
\begin{align} \label{eq:surface_density_gc}
\Sigma_{\rm GC}(R)
=
\Sigma_e
\exp\left[
-b_n
\left(
\left(\frac{R}{R_{\rm GC}^{\rm eff}}\right)^{1/n}
-1
\right)
\right],
\end{align}
where $\Sigma_e$ is the surface density at the effective radius $R_{\rm GC}^{\rm eff}$ and $n$ is the Sérsic index. The coefficient
\begin{align}
b_n
=
2n
-\frac{1}{3}
+\frac{4}{405\,n}
+\mathcal{O}(n^{-2})
\end{align}
is chosen such that $R_{\rm GC}^{\rm eff}$ encloses half of the total projected \gls{GC} population. Observationally, \gls{GC} systems are typically described by Sérsic indices in the range $n\simeq1$--$3$, with larger values generally found in massive elliptical galaxies~\citep{2022MNRAS.510.5725D,2025ApJS..276...34L}. For the effective radius we use the empirical scaling relation of~\citep{2018MNRAS.477.3869H},
\begin{align}
R_{\rm GC}^{\rm eff}
\simeq
22\,{\rm kpc}
\left(
\frac{M_{\rm h}}{10^{13}\,M_\odot}
\right)^{0.88}.
\end{align}
Finally, the normalization is fixed by requiring that the projected surface density integrates to the total number of \glspl{GC} in the halo,
\begin{align}
N_{\rm GC}
=
2\pi
\int_0^\infty
R\,\Sigma_{\rm GC}(R)\,{\rm d}R.
\end{align}
Evaluating the integral gives
\begin{align}
\Sigma_e
=
\frac{N_{\rm GC}}
{2\pi (R_{\rm GC}^{\rm eff})^2
\,n\,e^{b_n}\,
b_n^{-2n}\,
\Gamma(2n)},
\end{align}
where $\Gamma(x)$ denotes the Gamma function.

\section{Gravitational-wave strong lensing by  \glspl{GC}  with  \glspl{IMBH} }

\begin{figure*}[t]
    \centering
    \includegraphics[width=0.475\linewidth]{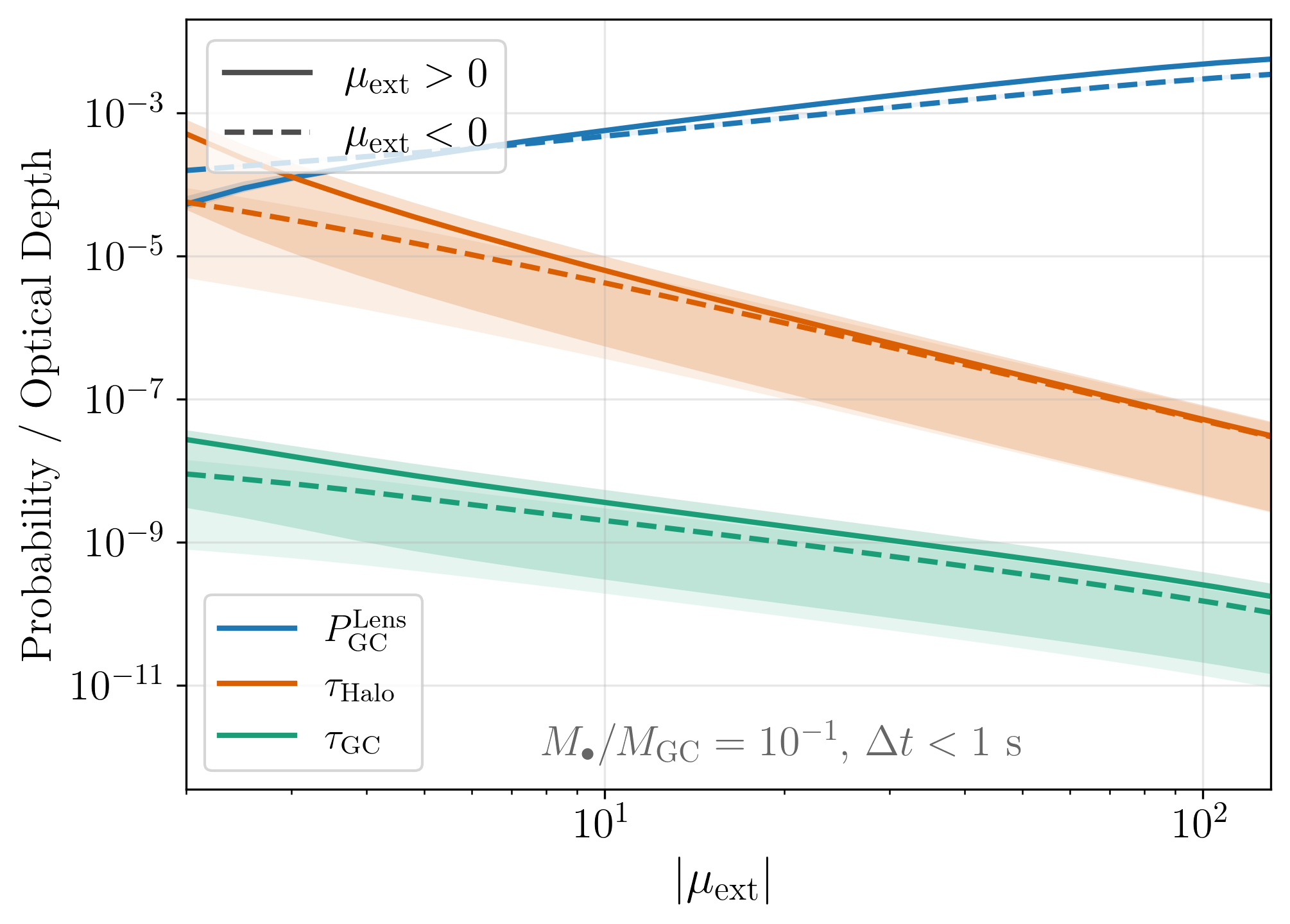}
    \includegraphics[width=0.475\linewidth]{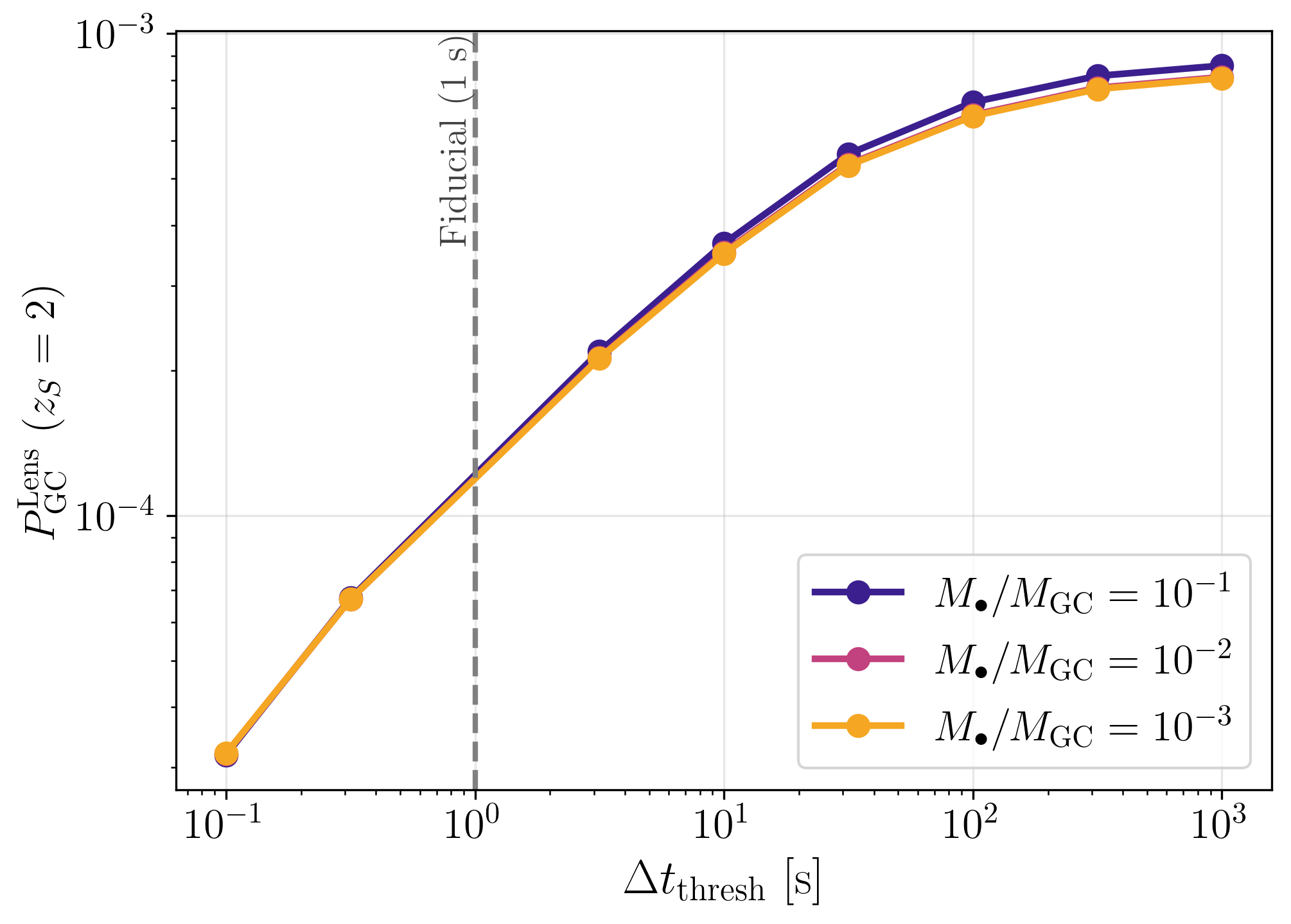}
    \caption{(Left) Conditional probability of lensing for $M_{\bullet}/M_\mathrm{GC} = 0.1$ shown against the magnification threshold. 
    The solid line corresponds to the probability for \glspl{GC} outside of the main halo critical curves, and the dashed lines correspond to those inside the critical curves. 
    Shaded regions show the extent of the probability over the source redshift range $z_{\rm S}\in[0.5,3.0]$. The individual optical depths of the \glspl{GC} and the main halos are provided for reference. 
    (Right) Conditional probability as a function of the time delay threshold for the two brightest images for different \gls{IMBH} masses. 
    Our fiducial threshold is 1 second.
    }
    \label{fig:tau_mass_ratio}
\end{figure*}

When a \gls{GW} is strongly lensed by a galactic halo, it produces repeated copies of the original signal. 
Each of these copies can be (de)magnified, delayed and phase shifted~\citep{1992grle.book.....S,Ezquiaga:2020gdt}. 
Because current ground-based \gls{GW} detectors are limited in their distance reach, the most likely first detections will have high magnifications, $\langle\mu\rangle\sim15$~\citep{Oguri:2018muv,Xu:2021bfn}.   
The magnification produced by the halo depends on the convergence $\kappa$ and shear $\gamma$, which we will refer as ``external'' to distinguish from the \gls{IMBH} and \gls{GC} contributions:
\begin{equation}
    \mu_{\rm ext} = \frac{1}{(1-\kappa_{\rm ext})^2 - \gamma_{\rm ext}^2}\,.
\end{equation}
The image positions with maximum magnification define the \emph{critical curves}, which are determined by the zeros of the denominator. 
Critical curves are mapped onto \emph{caustics} in the source plane. 
When a source crosses a caustic, two images are created. 

We model the main halo as a \gls{SIS} truncated at the virial radius:
\begin{equation}
    \rho(r)=\frac{\sigma_v^2}{2\pi G r^2}\,,\qquad r<r_v\,,
\end{equation}
where $\sigma_v$ is the velocity dispersion. 
The virial radius $r_v$ determines the extent of the halo, $r_{\rm h}$. 
The density at this radius, $\rho(r_{\rm h})$, can be related to the critical density of the Universe at a given redshift, $\rho(z)=\Delta_h \rho_c(z)$ where $\rho_c(z)=3H^2(z)/8\pi G$. 
By convention, we choose $\Delta_h = 200$. 
Using the virial theorem, $V^2=GM_{\rm h}/r_{\rm h}$, which relates the circular velocity at the virial radius $V$ with the halo mass $M_{\rm h}$, we can relate the velocity dispersion to the virial mass:
\begin{equation} \label{eq:sigma_M_virial}
    \sigma_v = \left(\frac{\pi \rho(r_v)}{6}\right)^{1/6}G^{1/2}M_v^{1/3}\,,
\end{equation}
where we have used that for a \gls{SIS}, $V=\sqrt{2}\sigma_v$. 
Inserting the density of the halo at the virial radius, the expression becomes
\begin{equation}
    \sigma_{\rm h}^3 = \frac{\sqrt{\Delta_h}}{4}GH(z)M_{\rm h}.
\end{equation}

The relevant scales for lensing are determined by the Einstein radius
\begin{equation}
    \theta_{\rm SIS}^2 = 4\pi\left(\frac{\sigma_v}{c}\right)^2 \frac{ D_{LS}}{D_S}\,,
\end{equation}
which, in addition to the velocity dispersion, depends on the angular diameter distances to the source $D_S$ and between the lens and the source $D_{LS}$. 
The \gls{SIS} is characterized by $\kappa=\gamma$, and therefore the critical curves are simply defined by the image positions at which $\kappa=1/2$. 
The external magnification is thus simply related to the convergence by
\begin{equation}
    \mu_{\rm ext}=\frac{1}{1-2\kappa_{\rm ext}}\,.
\end{equation}

In this model, two images of opposite parity are produced when the source is within the Einstein radius (labeled as $\pm$), and one otherwise. 
The source-plane cross-section for a magnification of one of the images to be larger than a given threshold, $|\mu_\pm|>\mu_0$, follows
\begin{equation} \label{eq:SIS_mu_cross_section}
    \hat\sigma(|\mu_\pm|>\mu_0)=\frac{1}{(\mu_0\mp1)^2}\pi \theta_{\rm SIS}^2\,.
\end{equation}
This circle in the source plane maps into an annulus in the image plane. 
The positive parity image is always outside the critical curve with a radial position constrained by
\begin{equation}
    \frac{\theta_+}{\theta_{\rm SIS}}<\frac{1}{1-1/\mu_0}\,,
\end{equation}
while the negative parity image is inside the critical curve 
\begin{equation}
    \frac{\theta_-}{\theta_{\rm SIS}}>\frac{1}{1+1/\mu_0}\,.
\end{equation}
We can map back the image radial position to the convergence:
\begin{equation} \label{eq:R_kappa_relation_SIS}
    |\theta|=\frac{\theta_{\rm SIS}}{2\kappa}\,.
\end{equation}
This allows us to know the magnification associated with each projected radial distance in the image plane. 

As mentioned before, we model also the \gls{GC} as an \gls{SIS}. 
Its mass, $M_{\rm GC}$, is related to its velocity dispersion by 
\begin{equation}
    \sigma_{\rm GC}^2=\frac{GM_{\rm GC}}{2r_v}\,,
\end{equation}
where the \gls{SIS} truncation scale is fixed to a virial radius twice the half-mass radius. 
The \gls{IMBH} is parametrized with a point-mass lens $M_\bullet$ located at the center of the cluster. 
Isolated point-mass lenses always produce two images. 

The \gls{GC} and \gls{IMBH} are embedded in the galactic halo, which is controlled by $\kappa_{\rm ext}$. 
The total lensing potential $\psi_{\rm tot}$ has then three contributions
\begin{equation}
    \psi_{\rm tot}=\psi_{\rm halo}(\kappa_{\rm ext}) + \psi_{\rm GC}(M_{\rm GC})+\psi_{\rm IMBH}(M_\bullet)\,.
\end{equation}
The presence of the external convergence and shear breaks the axisymmetry, modifying the caustic structure compared to the isolated case. 
A schematic representation of the setup is shown in Fig. \ref{fig:the_problem}.
Namely, spherically symmetric perturbers at $\kappa_{\rm ext} < 0.5$ produce the familiar astroid-shaped caustics found in elliptical lenses (cf. middle panel of Fig.~\ref{fig:the_problem}). 
These astroids are stretched along the shear axis as $\kappa_{\rm ext} \rightarrow 0.5$, until they pass the critical curves and become two disconnected mirrored caustics (right panel of Fig.~\ref{fig:the_problem}), each comprised of three cusps and three folds \citep{An:2006bq}. 
Each of these configurations will produce 4 images when a source is placed within their caustics. 
We use the composite lens potential presented in Appendix~\ref{app:lensmodel} to solve for the image properties, and validated results numerically against the widely used lensing code \texttt{lenstronomy}~\citep{Birrer:2018xgm, Birrer:2021wjl}.

To quantify the probability of lensing for a source at redshift $z_{\rm S}$, we compute the optical depth $\tau(z_{\rm S})$, which integrates the lensing cross section $\hat \sigma$ relative to the sky area ($4\pi$) for all potential lenses up to $z_{\rm S}$. 
In the case of strong lensing by the main halo, the probability of having a magnification larger than $\mu_0$ follows from Eq. (\ref{eq:SIS_mu_cross_section}) and is determined by
\begin{equation}
\begin{split}
    \tau_{\rm halo}(z_{\rm S}&)=\int_0^{z_{\rm S}}{\rm d}z_{\rm L}\int_{M_{\rm min}}^{M_{\rm max}} 
    {\rm d}M_{\rm h} \frac{{\rm d}n_{\rm h}}{{\rm d}M_{\rm h}}\frac{{\rm d}V_c}{{\rm d}z_{\rm L}} \frac{\hat\sigma_{\rm halo}(M_{\rm h},\mu_0)}{4\pi}\,,
\end{split}
\end{equation}
where ${\rm d}n_{\rm h}/{\rm d}M_{\rm h}$ is the number density of halos, ${\rm d}V_c/{\rm d}z=(1+z)^24\pi D_L^2/H(z)$ is the differential comoving volume which depends on the angular diameter distance to the lens $D_L$. 
We model the host halos with a Schechter function fitted to the observations of the Sloan Digital Sky Survey~\citep{Choi:2006qg}. 
They are integrated in the range $[10^{11},10^{14}]M_\odot$, which covers the whole velocity dispersion distribution peaking at 161 km/s. 

The optical depth for lensing by a \gls{GC} with an \gls{IMBH} is obtained by integrating the projected surface density (\ref{eq:surface_density_gc}) 
in the annulus defined by the threshold magnification of the main halo, $R^{\rm min/max}_{\mu_0}$. 
Note that the optical depth will be different for the even and odd parity images due to the structural difference of the caustics on either side of the critical curves. 
The result is
\begin{equation}
\begin{split}
    \tau_{\rm GC}(z_{\rm S}&|M_{\rm GC},M_\bullet,N_\bullet)=\int_0^{z_{\rm S}}{\rm d}z_{\rm L}\int {\rm d}M_{\rm h} \frac{{\rm d}n_{\rm h}}{{\rm d}M_{\rm h}}\frac{{\rm d}V_c}{{\rm d}z_{\rm L}} \times \\
    &\int_{R^{\rm min}_{\mu_0}}^{R^{\rm max}_{\mu_0}}{\rm d}R\,2\pi R\, \Sigma_{\rm GC}(R)\frac{\hat \sigma_{\rm GC}(R|M_{\rm GC},M_\bullet,N_\bullet)}{4\pi}\,,
\end{split}
\end{equation}
where $\hat \sigma_{\rm GC}(R|M_{\rm GC},M_\bullet,N_\bullet)$ is the cross section for lensing by the \gls{GC} with an \gls{IMBH} that depends on their masses and occupation fraction. 
Note that the dependence in the projected radius $R$ links directly to the external convergence and magnification via (\ref{eq:R_kappa_relation_SIS}). 
With the optical depths we can also compute the fraction of strongly lensed events by the main halo that would be lensed by the \gls{GC} and \gls{IMBH}. 
The conditional probability is
\begin{equation}
    P^{\rm Lens}_{\rm GC}(\tau_{\rm GC}(z_{\rm S}) |\tau_{\rm halo}(z_{\rm S}) ) \equiv \frac{\tau_{\rm GC}(z_{\rm S})}{\tau_{\rm halo}(z_{\rm S})}
\end{equation}
The fractional rate of lensing in strongly lensed \glspl{GW}  will be our main target.

The \gls{GC} cross section for lensing is defined as the area in the source plane in which images with time delays of $\Dt < 1$ second, as well as a relative magnification of $\mu_{\rm rel} > 0.1$ are produced. The first criterion leads to overlapping signals characteristic of lensing signatures in lensed \gls{LVK} band sources. 
The latter ensures that the second image can be distinguished as a lensed image associated with the same event, as indicated in large-scale parameter estimation studies~\citep{Chan:2025pdf}.

\section{Expected rates}
\label{sec:expected_rates}

We compute the expected fractional rate of lens in strongly lensed \glspl{GW}, $P^{\rm Lens}_{\rm GC}(\tau_{\rm GC}(z_{\rm S}) |\tau_{\rm halo}(z_{\rm S}) )$. 
We begin by exploring how this rate changes as a function of the minimum external magnification for a fixed mass ratio $M_\bullet/M_{\rm GC}=0.1$. Our fiducial source redshift is 2 (solid and dashed lines), but we indicate with a band other source redshifts from 0.5 to 3. 
We compute this separately for positive (solid) and negative (dashed) external magnifications.  
The results are presented in the left panel of Fig. \ref{fig:tau_mass_ratio}. 
The fractional rate is strongly dependent on the minimum external magnification, varying from $10^{-4}$ to $10^{-2}$ for external magnification thresholds from 2 to 100. 
Larger external magnifications are intrinsically more rare, as shown by comparing the optical depths of the main halo and \glspl{GC}. 
They however hold universal observational predictions~\citep{Lo:2024wqm}, even in the diffraction regime~\citep{Ezquiaga:2025gkd}, and could be more frequent in cluster-scale lenses~\citep{Vujeva:2025kko}. 
We also find that the relative lensing rate is less sensitive to the source redshift than the optical depths themselves. 

We then investigate the impact of the time delay threshold between the two brightest images in the fractional rate of lens. 
As shown in the right panel of Fig. \ref{fig:tau_mass_ratio}, the rate increases until it plateaus at the value in which only the relative magnification threshold is imposed, which is set at $\sim10^{-3}$. 
This maximum value is reached for time delays of tens of minutes, which would not produce overlapping images for \glspl{BBH} in current ground-based GW detectors. 
Our fiducial threshold of 1 second leads to about an order of magnitude lower relative rate. 
We observe that $P^{\rm Lens}_{\rm GC}$ does not depend strongly on the mass ratio between the \gls{IMBH} and \gls{GC}. 
We notice that almost all configurations that produce a sub-second image pair already have $\mu_{\rm rel}$ comfortably above 0.1. 
Their rate only starts to reduce above 0.5. 

\begin{figure}[t]
    \centering
    \includegraphics[width=\columnwidth]{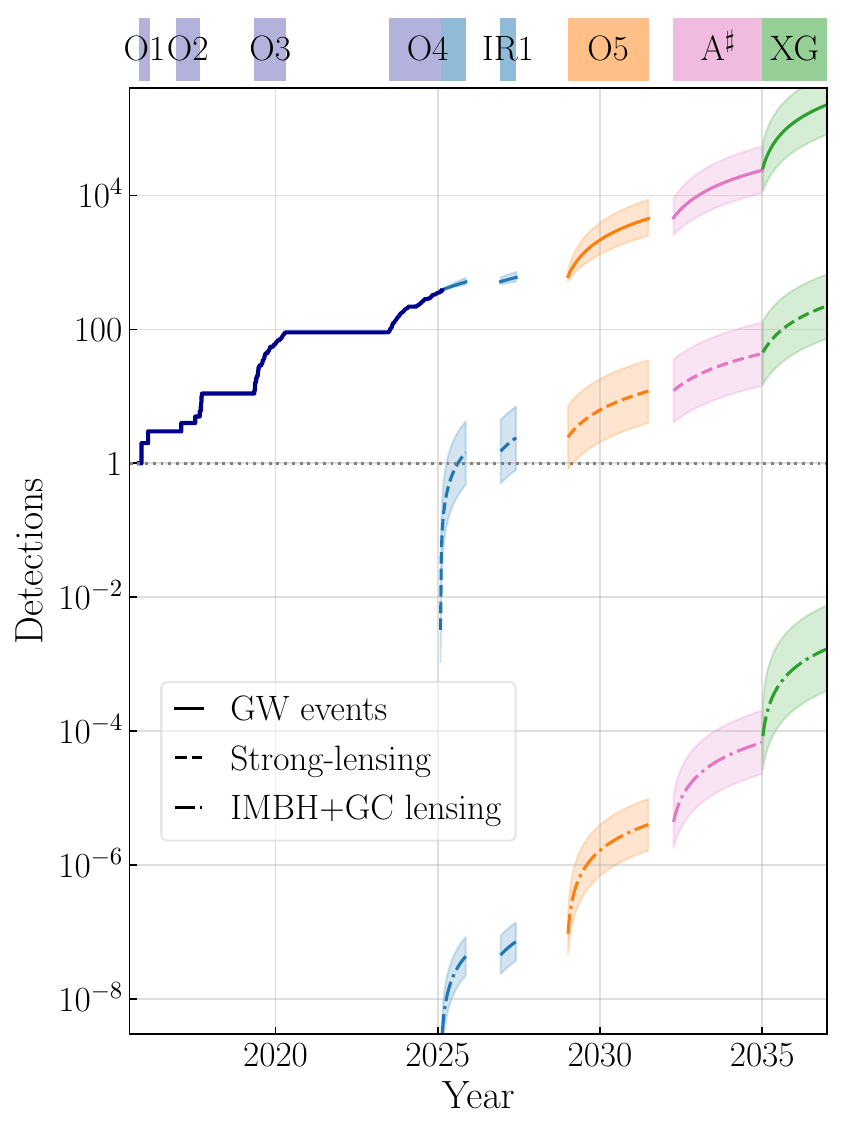}
    \caption{Projection for the number of lensed events as a function of time for our fiducial model with an \gls{IMBH} occupation fraction in \glspl{GC} of 1. 
    The number of events lensed by the compound \gls{IMBH} and \gls{GC} configurations are found to always lie well below 1 detection. 
    For reference, we also show the number of expected \glspl{GW} and strongly lensed \glspl{GW}. 
    Projections include uncertainties in the observed \gls{BBH} merger rate. 
    }
    \label{fig:lensing_rates}
\end{figure}

We perform robustness tests to quantify how the quoted rates depend on the underlying modeling assumptions. 
In particular, we vary the assumptions about the radial distribution of  \glspl{GC} , changing the Sérsic index and considering a Hubble profile. 
We find that the rates only change by factors of 0.7--3. 
Varying the \gls{GC} virial radius between 4 pc and 8 pc changes the rates by $\sim20\%$. Additionally, we consider the effects of isolated \glspl{GC} (or \glspl{GC} at very large distances from the host galaxy). Due to the vast majority of the total optical depth coming from \glspl{GC} near the critical curves, the rate for lensing by such systems is far lower than the rates reported for the fiducial examples in this work.

Finally, we translate our optical depth calculations into expected observations. 
To that matter, we model a population of \gls{BBH} sources consistent with the inferred population from the latest \gls{GW} transient catalog~\citep{LIGOScientific:2026ctl}. 
The most important uncertainty for this forecast is the intrinsic rate itself, and for that reason we project results within its 90\% credible interval. 
We define a lensed detection as one in which at least two of the repeated chirps are above a signal-to-noise ratio of 8. 
We take this simplified and optimistic detection criterion to estimate the number of detectable lensed \glspl{GW}. 
The results are shown in Fig. \ref{fig:lensing_rates} as a function of time, for different future observing scenarios -- see Appendix \ref{sec:gw_observing_scenarios} for details. 
The figure assumes that no lensing detections have been made until GWTC-4.0~\citep{GWTC4_lensing}, and projects afterwards. 
While strongly lensed \glspl{GW}  by galaxy-scale halos are expected to be observed in the very near future, we see that even with next-generation (XG) detectors \glspl{GW}  lensed by  \glspl{IMBH}  in  \glspl{GC}  are far out of reach. 
Although we focus on ground-based detectors, our results extend to future \gls{GW} space antennas such as LISA~\citep{LISACosmologyWorkingGroup:2022jok}, where $\sim0.1$--100 strongly lensed massive \gls{BBH} detections are expected over the 4 year mission~\citep{Gutierrez:2025ymd,Sun:2026lrn}. 

\section{Implications}

The suppressed probability of \gls{GW} lensing by  \glspl{IMBH}  and \glspl{GC} has important implications for the search of lensed \glspl{GW} more broadly. 
First and foremost, it affects directly the interpretation of candidate lensed events such as GW231123~\citep{GW231123}, whose inferred lens mass is $\sim1000M_\odot$~\citep{GWTC4_lensing}. 
To make this explicit, we exploit the calculations done in the previous section to compute the Bayes factor and the odds ratio, which compare two competing hypothesis (lensed vs. nonlensed) taking into account how well they describe the data and their (astrophysical) prior odds. 
The details can be found in Appendix \ref{sec:Bayes_factor_and_odds_ratio_calculation}.

\begin{figure}[t]
    \centering
    \includegraphics[width=\columnwidth]{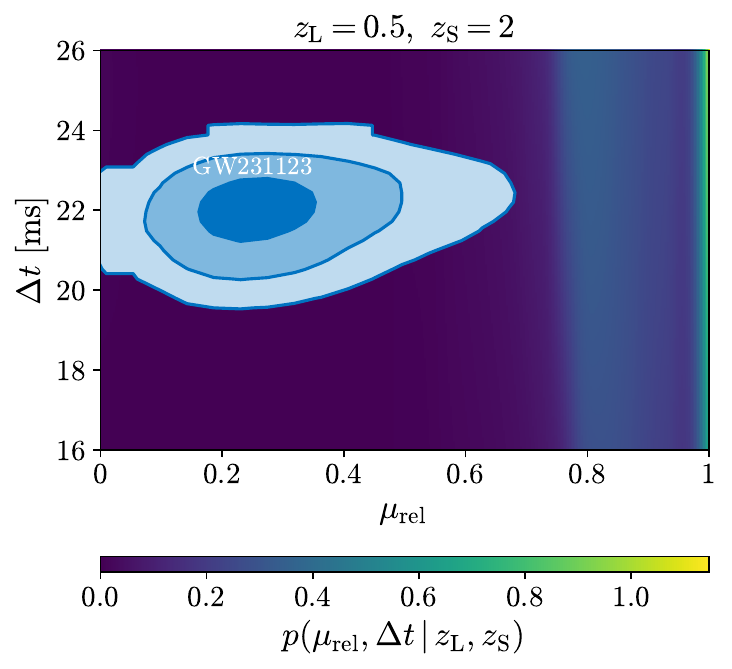}
    \caption{Inconsistency of the inferred lensing parameters $\mur, \Dt$ from GW231123 with the prediction from our lensing by \gls{IMBH}+\gls{GC} setup. The probability density $p(\mur{}, \Dt{}|\zl{}, \zs{})$ from our simulations, marginalized over the mass ratio $M_{\bullet}/M_\mathrm{GC}$, is computed at some fiducial lens and source redshift, $\zl{} = 0.5$ and $\zs{} = 2$, respectively.}
    \label{fig:GW231123_inconsistency}
\end{figure}

Given the Bayes factor of $\BFValue{}$ and odds ratio of $\ORValue{}$, we find 
that although this event has the highest support for lensing of any observed \gls{GW} event,  
imposing an astrophysical distribution of candidate lenses required to produce the inferred lens parameters further reinforces that this event is unlikely to be lensed by the family of embedded lenses studied in this work. This is further illustrated in Fig.~\ref{fig:GW231123_inconsistency}. We find this consistent with the astrophysical-prior analysis of \cite{Cheung:2026pky}, which disfavors a lensing interpretation of GW231123 by an isolated lens.
The majority of the support for the posterior distribution of the lensing parameter $\mur$ and $\Dt$ inferred from GW231123 lies in the region where our \gls{IMBH}+\gls{GC} lensing setup predicts to be rare, as shown by the probability density $p(\mur{}, \Dt{}|\zl{}, \zs{})$ marginalized over the mass ratio $M_{\bullet}/M_\mathrm{GC}$ at some fiducial lens and source redshift, $\zl{} = 0.5$ and $\zs{} = 2$, respectively.

However, the low likelihood of \glspl{IMBH} and \glspl{GC} 
producing 
new images or waveform distortions 
offers in return a particularly clean probe into new physics that \glspl{GW} are uniquely suited to probe. 
For instance, strongly lensed \glspl{GW} are also affected by dark matter substructures~\citep{Vujeva:2025nwg, Li:2026dai}. 
Their rate depends crucially on the subhalo concentration, which varies across different dark matter scenarios, and can already reach 1--10\% in cold dark matter models for a magnification of $\mu_{\mathrm{max}} > 25$~\citep{Vujeva:2025nwg}, much higher than the most optimistic \gls{IMBH} and \gls{GC} rate. 
This means that lensed \glspl{GW} could be an incredibly clean probe of dark matter substructures at the smallest scales. 
This is critical due to most candidate dark matter models finding their largest discrepancies at masses below $10^6 M_\odot$, precisely the range explored in this work.

Additionally, \glspl{PBH} formed in the early Universe can ostensibly produce black holes in the mass ranges relevant for lensing of \gls{LVK} sources~\citep{Sasaki:2018dmp}. 
\glspl{PBH} are constrained by many different observations at different levels and mass ranges~\citep{Carr:2020gox}, with \glspl{GW} lensing setting bounds only at the level of $\sim50\%$~\citep{LIGOScientific:2023bwz}, which could be lower when taking into account that highly distorted lensed signals could be missed by current matched-filtering pipelines~\citep{Chan:2024qmb}. 
We find that uniformly distributed \glspl{PBH} would lead to a number of lensed detections larger than the projected rates for  \glspl{IMBH}  in Fig. \ref{fig:lensing_rates} for a dark matter fraction larger than $\sim10^{-6}$. 
Therefore, \gls{GW} observations could be useful probes of primordial physics.

A noteworthy caveat to this are the compounding effects of stellar fields near critical curves, which have been of particular interest in recent years \citep{Diego:2019lcd,Mishra:2021xzz,Meena:2022unp,Shan:2022xfx,Meena:2023qdq, Zumalacarregui:2026uqs}. 
While \cite{Zumalacarregui:2026uqs} has 
found there to be non-negligible stochastic statistical deviations in the amplification function caused by stellar fields, these small deviations will be both qualitatively and quantitatively different from the more dramatic effects on a per-event basis caused by these more massive perturbers, and as such should not produce large effects until their densities resemble those of  \glspl{GC}   themselves. 
In this limit, the rate should approach the one found in this work, with qualitative differences in the lensed signal caused by the granularity of the lens (particularly near the critical curves of the main halo).

In summary, the likelihood of \gls{GW} lensing by \glspl{GC} and \glspl{IMBH} is inherently low. This low rate is itself an advantage: it leaves a regime in which searches for dark matter substructure are statistically unlikely to be confused with lensing by known astrophysical potentials.

\begin{acknowledgments}
We are grateful to Mark Cheung, Miguel Zumalacarregui, and Sreekanth Harikumar for their helpful comments and suggestions.
This work was supported by Research Grants No.~VIL37766 and No.~VIL53101 from Villum Fonden and the DNRF Chair Program Grant No.~DNRF162 by the Danish National Research Foundation.
This work has received funding from the European Union's Horizon 2020 research and innovation programme under the Marie Sklodowska-Curie Grant Agreement No.~101131233.  
J.~M.~E is also supported by the Marie Sklodowska-Curie Grant Agreement No.~847523 INTERACTIONS.
L.Z. is supported by the European Union’s Horizon 2024 research and innovation program under the Marie
Sklodowska-Curie grant agreement No. 101208914.
The Center of Gravity is a Center of Excellence funded by the Danish National Research Foundation under grant no.~DNRF184.
The Tycho supercomputer hosted at the SCIENCE HPC center at the University of Copenhagen was used for supporting this work.

This research has made use of data or software obtained from the Gravitational Wave Open Science Center (gwosc.org), a service of the LIGO Scientific Collaboration, the Virgo Collaboration, and KAGRA. This material is based upon work supported by NSF's LIGO Laboratory which is a major facility fully funded by the National Science Foundation, as well as the Science and Technology Facilities Council (STFC) of the United Kingdom, the Max-Planck-Society (MPS), and the State of Niedersachsen/Germany for support of the construction of Advanced LIGO and construction and operation of the GEO600 detector. Additional support for Advanced LIGO was provided by the Australian Research Council. Virgo is funded, through the European Gravitational Observatory (EGO), by the French Centre National de Recherche Scientifique (CNRS), the Italian Istituto Nazionale di Fisica Nucleare (INFN) and the Dutch Nikhef, with contributions by institutions from Belgium, Germany, Greece, Hungary, Ireland, Japan, Monaco, Poland, Portugal, Spain. KAGRA is supported by Ministry of Education, Culture, Sports, Science and Technology (MEXT), Japan Society for the Promotion of Science (JSPS) in Japan; National Research Foundation (NRF) and Ministry of Science and ICT (MSIT) in Korea; Academia Sinica (AS) and National Science and Technology Council (NSTC) in Taiwan.
\end{acknowledgments}

\appendix
\section{Composite Lens Model}
\label{app:lensmodel}

In this section we outline the details of our composite lens model comprised of an SIS and point-mass lens (centered at the same point in the lens plane), in the presence of external shear and convergence. While the image properties of such a composite lens system can be computed directly with existing lensing codes such as \texttt{lenstronomy}~\citep{Birrer:2018xgm, Birrer:2021wjl}, the heavy distortion of the caustics caused by placing the \gls{GC} near the critical curves of the main halo require a large simulation area to capture their extent, with a very high spatial resolution in order to capture its features. 

To lessen the cost of this brute force approach, we calculate the arrival times and magnification factors (and ratios) explicitly using the total lensing potential of the SIS + point-mass + external shear with the following form:

\begin{equation}
\begin{split}
\psi_{\mathrm{tot}}
    ={}& \frac{1}{2}\gamma_{\mathrm{ext}}(\theta_\mathrm{x}^2-\theta_\mathrm{y}^2)
        + \sqrt{\theta_\mathrm{x}^2+\theta_\mathrm{y}^2} \\
     &+ \frac{1}{2}\kappa_{\mathrm{ext}}(\theta_\mathrm{x}^2+\theta_\mathrm{y}^2)
        + f_{\mathrm{mass}}\log\!\left(\sqrt{\theta_\mathrm{x}^2+\theta_\mathrm{y}^2}\right),
\end{split}
\end{equation}
where $\vec\theta = (\theta_\mathrm{x},\theta_\mathrm{y})$ are the image plane coordinates, and $\vec\beta = (\beta_\mathrm{x},\beta_\mathrm{y})$ are the source plane coordinates. 
The image positions $\vec\theta_i$ of a source at $\vec\beta$ are the solutions of the
lens equation $\vec\beta = \vec\theta - \vec\nabla\psi_{\mathrm{tot}}$, which for this
potential takes the component form
\begin{equation}
\begin{split}
\beta_\mathrm{x}
={}& \theta_\mathrm{x}
\left[
    1-\kappa_{\mathrm{ext}}-\gamma_{\mathrm{ext}}
    - \frac{1}{\sqrt{\theta_\mathrm{x}^2+\theta_\mathrm{y}^2}}
    - \frac{f_{\mathrm{mass}}}{\theta_\mathrm{x}^2+\theta_\mathrm{y}^2}
\right], \\
\beta_\mathrm{y}
={}& \theta_\mathrm{y}
\left[
    1-\kappa_{\mathrm{ext}}+\gamma_{\mathrm{ext}}
    - \frac{1}{\sqrt{\theta_\mathrm{x}^2+\theta_\mathrm{y}^2}}
    - \frac{f_{\mathrm{mass}}}{\theta_\mathrm{x}^2+\theta_\mathrm{y}^2}
\right],
\end{split}
\end{equation}
equivalently the stationarity condition $\vec\nabla_{\vec\theta}\,\phi_{\mathrm{tot}}=0$
of the Fermat potential. 

The arrival-time difference between images $i$ and $j$ is then
\begin{equation}
\Delta t_{ij}
= T\left[\phi_{\mathrm{tot}}(\vec\theta_i)-\phi_{\mathrm{tot}}(\vec\theta_j)\right],
\end{equation}

\begin{equation}
\qquad
T=\frac{4GM_{E}\,(1+z_{L})}{c^{3}}
\simeq 1.97\,\mathrm{s}\, 
\left(\frac{M_{E}}{10^{5}\,M_{\odot}}\right)\left(1+z_{L}\right),
\end{equation}
where $M_E$ is the projected SIS mass within $\theta_{\mathrm{SIS}}$

Finally, the magnification factor of each image can be computed by taking the determinant of the lensing Jacobian, which gives
\begin{equation}
\begin{split}
\mu_{\mathrm{tot}}^{-1}
={}& -\gamma_{\mathrm{ext}}^2 \\
&+ 
    \frac{2\gamma_{\mathrm{ext}}(\theta_\mathrm{x}^2-\theta_\mathrm{y}^2)}{{
    \theta_\mathrm{x}^2+\theta_\mathrm{y}^2
}}
    \left(
        \dfrac{f_{\mathrm{mass}}}{\theta_\mathrm{x}^2+\theta_\mathrm{y}^2}
        + \dfrac{1}{2\sqrt{\theta_\mathrm{x}^2+\theta_\mathrm{y}^2}}
    \right)
 \\
&+ \left(
    1-\kappa_{\mathrm{ext}}
    - \frac{1}{2\sqrt{\theta_\mathrm{x}^2+\theta_\mathrm{y}^2}}
\right)^2 \\
&- \left(
    \frac{f_{\mathrm{mass}}}{\theta_\mathrm{x}^2+\theta_\mathrm{y}^2}
    + \frac{1}{2\sqrt{\theta_\mathrm{x}^2+\theta_\mathrm{y}^2}}
\right)^2 .
\end{split}
\end{equation}

These results are found to be in good agreement with the numerical implementation of an identical setup in \texttt{lenstronomy}. 
We also verify that in the limits in which either the point-mass lens or SIS components of the lens are removed, we recover the known properties of the point-mass + $\gamma_{\rm ext}$~\citep{An:2006bq} and SIS + $\gamma_{\rm ext}$~\citep{Finch:2002ww}.

\section{Gravitational-wave observing scenarios}
\label{sec:gw_observing_scenarios}

In order to forecast the number of lensed detections, cf. Fig. \ref{fig:lensing_rates}, we follow the \gls{GW} observing scenarios described in~\cite{KAGRA:2013rdx}.\footnote{Such plans are regularly updated at \href{https://observing.docs.ligo.org/plan/}{https://observing.docs.ligo.org/plan/}. We use the July 15, 2026 update.} 
In particular, we consider that the four already completed runs (O1-O4) are followed by a 6 months intermediate run (IR1) with equal sensitivity to O4. 
After that, plans are more uncertain, but we assume the currently projected fifth observing run (O5) at advanced LIGO design sensitivity~\citep{sensitivity_curves_ligo}. 
Next to that is A$^\sharp$~\citep{T2200287}, which exploits LIGO sites but with upgraded detectors. 
Finally, we compute rates for next-generation (XG) ground-based detectors such as Cosmic Explorer~\citep{Evans:2021gyd} and Einstein Telescope~\citep{et}. 

\section{Bayes factor and Odds ratio calculation}
\label{sec:Bayes_factor_and_odds_ratio_calculation}
Following \cite{Lo:2021nae}, we can compute the Bayes factor and the odds ratio comparing the two hypotheses for GW231123, namely the hypothesis that it is not lensed/a vanilla \gls{BBH} (denoted by $\HNL{}$) and the alternative hypothesis that it is a \gls{BBH} lensed by \gls{IMBH}+\gls{GC} (denoted by $\HL{}$).

The Bayes factor $\BayesFactor$ is given by the ratio of the data likelihood under the two hypotheses, respectively. Mathematically, it can be written as
\begin{equation}
\label{eq:BF_def}
    \BayesFactor \equiv \frac{P(D|\HL{})}{P(D|\HNL{})},
\end{equation}
where we ignore the differences in the selection effects for the two hypotheses here (for details, see \cite{Chan:2024qmb}).
The data likelihood $P(D|\HNL{})$ under the non-lensed hypothesis can be obtained by reweighting the output from the \gls{PE} done in \cite{Chan:2025kyu} to account for the differences between the sampling prior used and the \gls{GW} source population model $p_{\rm BBH-pop}$ adopted here, which is taken to be the default \gls{BBH} population model in \cite{GWTC4_pop}.

As for the data likelihood $P(D|\HL{})$ under the lensed hypothesis, it involves slightly more work since we need to marginalize over the (unknown) lens redshift and true source redshift. Specifically, we factorize the probability distributions as
\begin{equation}
\label{eq:data_likelihood_lensed}
    P(D|\HL{}) = \int \mathcal{L}(\theta|\HL{}) p(\theta | \zl{}, \zs{}, \HL{})p(\zl{}|\zs{}, \HL{}) p(\zl{}|\zs{}, \HL{})p(\zs{})\, d\theta d\zs{}d\zl{},
\end{equation}
where $\mathcal{L}(\theta|\HL{})$ is the likelihood given as a function of the waveform parameter $\theta$, and $p(\theta | \zl{}, \zs{}, \HL{})$ is the population-informed prior for those parameters. We further factorize the population-informed prior as
\begin{equation}
    p(\theta | \zl{}, \zs{}, \HL{}) = p(\mur{}, \Dt{}|\zl{}, \zs{})p_{\rm BBH-pop}(\theta \setminus \left\{ \mur{}, \Dt \right\}),
\end{equation}
where $p(\mur{}, \Dt{}|\zl{}, \zs{})$ comes from our simulations, marginalized over the mass ratio of the \gls{IMBH} to the \gls{GC}.
In practice, we approximate the integration over $\theta$ by a Monte Carlo integral over the posterior samples of $\theta$ from \gls{PE} done in \cite{Chan:2025kyu} and evaluate Eq.~\ref{eq:data_likelihood_lensed} using stochastic sampling \citep{Lo:2021nae}.
All these gives us the value of the Bayes factor as $\log_{10} \BayesFactor{} = \BFLogTenValue{}$.

The odds ratio $\OddsRatio$ is then the Bayes factor $\BayesFactor$ multiplied by the prior odds, which we set based on the average expected rate of \gls{GW} detection and lensing by \gls{IMBH} + \gls{GC} (see the calculations done in Sec.~\ref{sec:expected_rates}) as
\begin{equation}
\label{eq:prior_odds}
    \PriorOdds{} \equiv \frac{P(\HL{})}{P(\HNL{})} = \frac{7.8 \times 10^{-8}}{168} \approx 4.64 \times 10^{-10}.
\end{equation}
This gives us the value of the odds ratio $\OddsRatio{} = \ORValue{}$.

\bibliography{references}{}
\bibliographystyle{aasjournal}

\end{document}

%% file: new_commands.tex
\newcommand{\ba}{\begin{eqnarray}}
\newcommand{\ea}{\end{eqnarray}}
\newcommand{\be}{\begin{equation}}
\newcommand{\ee}{\end{equation}}

\newcommand{\R}{\mathcal{R}}

\newcommand{\zl}{z_{\text{L}}}
\newcommand{\zs}{z_{\text{S}}}

\definecolor{grey}{rgb}{0.4,0.4,0.4}
\definecolor{dullmagenta}{rgb}{0.4,0,0.4}
\definecolor{darkblue}{rgb}{0,0,0.4}
\definecolor{midblue}{rgb}{0,0,0.5}
\definecolor{midred}{rgb}{0.5,0,0}
\definecolor{orange}{rgb}{1,0.5,0}
\definecolor{lightbrown}{rgb}{0.75,0.5,0.25}
\definecolor{tan}{cmyk}{0.14,0.42,0.56,0}
\definecolor{djunglegreen}{cmyk}{0.99,0,0.52,0}
\definecolor{lightgreen}{rgb}{0,1,0}
\definecolor{olivegreen}{cmyk}{0.64,0,0.95,0.40}
\definecolor{midgreen}{rgb}{0.0,0.675,0.0}
\definecolor{darkgreen}{rgb}{0,0.5,0}
\definecolor{oxblood}{rgb}{0.5333, 0.0314, 0.0314}

\newcommand{\mur}{\mu_{\rm rel}}

\newcommand{\Dt}{\Delta t}

\newcommand{\HL}{\mathcal{H}_{\rm L}}
\newcommand{\HNL}{\mathcal{H}_{\rm NL}}
\newcommand{\PriorOdds}{\mathcal{P}^{\HL{}}_{\HNL{}}}
\newcommand{\BayesFactor}{\mathcal{B}^{\HL{}}_{\HNL{}}}
\newcommand{\OddsRatio}{\mathcal{O}^{\HL{}}_{\HNL{}}}

\newcommand{\BFLogTenValue}{-2.22}
\newcommand{\BFValue}{1/166}
\newcommand{\ORValue}{2.80 \times 10^{-12}}